# Re-entrance of Gapless Quantum Spin Liquids Observed in a Newly Synthesized Spin-1/2 Kagome Antiferromagnet $ZnCu_3(OH)_6SO_4$

Yuesheng Li,[1] Bingying Pan,[2] Shiyan Li,[2] Wei Tong,[3] Langsheng Ling,[3] Zhaorong Yang,[4] Junfeng Wang,[5] Zhongjun Chen,[6] Zhonghua Wu,[6] and Qingming Zhang[1*]

[1]Department of Physics, Renmin University of China, Beijing 100872, P. R. China

[2]State Key Laboratory of Surface Physics, Department of Physics, and Laboratory of Advanced Materials, Fudan University, Shanghai 200433, P. R. China

[3]High Magnetic Field Laboratory, Hefei Institutes of Physical Science, Chinese Academy of Sciences, Hefei 230031, P. R. China

[4]Key Laboratory of Materials Physics, Institute of Solid State Physics, Chinese Academy of Sciences, Hefei 230031, P. R. China

[5]Wuhan National high magnetic field center, Wuhan 430074, P. R. China

[6]Institute of High Energy Physics, Chinese Academy of Science, Beijing 100049, P. R. China

---

[*] qmzhang@ruc.edu.cn

**Quantum spin liquid (QSL) is a novel state of matter with exotic excitations and was theoretically predicted to be realized most possibly in an S=1/2 kagome antiferromagnet. Experimentally searching for the candidate materials is a big challenge in condensed matter physics and only two such candidates were reported so far. Here we report the successful synthesis of a new spin-1/2 kagome antiferromagnet $ZnCu_3(OH)_6SO_4$. No magnetic ordering is observed down to 50 mK, despite a moderately high Weiss temperature of $\theta_W \sim -79$ K. It strongly suggests that the material is a new QSL candidate. Most interestingly, the magnetic specific heat clearly exhibits linear behaviors in two low-temperature regions. Both behaviors exactly correspond to two temperature-independent susceptibilities. These consistently reveal a novel re-entrance phenomenon of gapless QSL state at the lowest temperatures. The findings provide new insights into QSL ground and excited states and will inspire new theoretical and experimental studies.**

Spin frustration describes a situation where spins cannot find an overall orientation configuration to simultaneously satisfy all of the nearest neighbor interactions[1]. In a strongly frustrated spin system, typical long-range magnetic ordering is prohibited and macroscopically degenerate ground states are produced. The ground state and low-energy excitations of a frustrated spin system have captured the greatest interests in condensed matter physics for several decades since the long-range resonating valence bond (RVB) state was proposed by Anderson, not only due to their relation to high temperature superconductivity, but also due to their remarkable collective phenomena[2-5]. The spin frustration was considered to be strongest in a spin-1/2 kagome Heisenberg antiferromagnet (KHA), which received extensive theoretical studies and many kinds of novel quantum spin liquid (QSL) ground states were proposed[6-11]. On the experimental side, only a few compounds were discovered and reported to show QSL features so far. Experimentally they exhibit Fermi-liquid-like low-energy excitations such as linear temperature dependence of specific heat and/or temperature-independent susceptibility, which were considered to arise from spinons, the fractional particle excitations in a gapless spin liquid[12-16].

Currently, one of the key issues in the field is to search for new QSL candidates. Several key factors are essential in searching promising QSL candidates. Magnetic ions with small spins are the first one, due to their strong quantum spin fluctuations at low temperatures. The others are a triangle-based lattice, a lower magnetic dimension and a smaller coordination number, etc. These put strong constraints on potential materials. Spin-1/2 KHA-type compounds meet the strict requirements. So far, only two spin-1/2 kagome antiferromagnets were discovered. One is the vanadium-based organic compound with S=1 interlayer ions[15]. The other one is herbertsmithite $ZnCu_3(OH)_6Cl_2$[17-21]. The 2D nature of spin lattices in the latter compound may be affected by an effective magnetic coupling between kagome planes, which is induced by anti-site disorder due to a small mixing of magnetic $Cu^{2+}$ and non-magnetic $Zn^{2+}$ between kagome and interlayer sites. And Dzyaloshinskii-Moriya (DM) interaction, which is a measure of spin-orbit coupling, was found to play a role in the compound[22,23].

$ZnCu_3(OH)_6Cl_2$ was synthesized by substituting a quarter of $Cu^{2+}$ in clinoatacamite $Cu_4(OH)_6Cl_2$, whose Neel temperature is ~ 7 K, with $Zn^{2+}$. Due to the different site preferences of $Cu^{2+}$ and $Zn^{2+}$, $Zn^{2+}$ ions enter into interlayer sites and kagome planes are dominated by $Cu^{2+}$ [24-26]. Inspired by the synthesis, we applied the similar method to a recently synthesized brochantite $Cu_4(OH)_6SO_4$, which has four symmetrically inequivalent Cu sites forming corrugated distorted triangular 2D planes and shows a long-range Neel ordering at 7.5 K[27]. After one of four symmetrical Cu sites (Cu4) in $Cu_4(OH)_6SO_4$ is replaced by $Zn^{2+}$, Cu-1/2 kagome antiferromagnet $ZnCu_3(OH)_6SO_4$ with well magnetically separated corrugated 2D planes is obtained. As the magnetic coordination number of 6 in triangle antiferromagnet $Cu_4(OH)_6SO_4$ is reduced to 4 in kagome $ZnCu_3(OH)_6SO_4$, the constraints on orientations of spins are much relaxed. Hence obvious suppression of long-range Neel ordering and much stronger spin frustrations can be expected. Different from herbertsmithite, both $Cu^{2+}$ and $Zn^{2+}$ ions in the new compound are located on the corrugated kagome planes.

Magnetization and heat capacity measurements reveal some exciting features in the compound. No magnetic ordering was observed even down to 50 mK. The $\theta_W \sim$ -79 K, gives a degree of spin frustration higher than 1580. A linear specific heat and a temperature-independent susceptibility, are observed from 6 ~ 15 K. The slope of linear specific heat and temperature-independent susceptibility are proportional to density of low-energy states and attributed to possible spinon excitations[5]. The derived Wilson ratio (~1.9) is much less than that of inorganic spin liquid (SL) materials, but close to that of organic SL ones[5,13], implying a small spin-orbit coupling in the new compound.

Most surprisingly, as cooling down to 0.6 K, the spin system clearly re-enters into a gapless QSL state with a much raised density of low-energy states, and maintains at least down to 50 mK. The observations seem compatible with the RVB QSL with a "pseudo-Fermi surface"[3,4]. The novel re-entrance of QSL has never been revealed before. It offers a completely new insight into QSL and its low-energy excitations.

Synchrotron X-ray powder diffraction and Rietveld refinement for $ZnCu_3(OH)_6SO_4$ are shown in Fig. 1(a). No additional peak is observed, which implies that impurity phases are negligible in the sample. $ZnCu_3(OH)_6SO_4$ has a monoclinic structure with space group P $2_1$/a, a = 13.0606(12) Å, b = 9.8697(10) Å, c = 6.0882(6) Å, and β = 103.6071(24)°, similar to that of $Cu_4(OH)_6SO_4$[27]. The difference is that one of the four different Cu sites (Cu4) in $Cu_4(OH)_6SO_4$ has been dominantly occupied by Zn (See Supporting Information), as synchrotron resonant

X-ray diffraction and combined Rietveld refinements revealed[28]. The crystal structure and ∠CuOCu bond angles are shown in Fig. 1(b). The ∠CuOCu bond angles suggest an antiferromagnetic (AF) coupling-dominated configuration[29,30]. In Fig. 1(b) and (c), there are two non-equivalent chains along c-axis. Edge-sharing copper octahedrons ($CuO_6$) lines up along chain A, and chain B consists of edge-sharing zinc planes ($ZnO_4$) and copper octahedrons ($CuO_6$). The chains are AB-stacked along b-axis by corners or edges to form 6% Cu-Cu bond-distorted corrugated kagome planes (Fig. 1(d)). The corrugated Cu-1/2 kagome planes are well magnetically separated by non-magnetic $SO_4$ tetrahedrons (Fig.1(e)). The spin triangle, a building block of spin lattices, consists of three non-equivalent nearest $Cu^{2+}$ ions, implying an anisotropic spin interaction. The site disorder between $Cu^{2+}$ and $Zn^{2+}$ is ~ 8% as revealed by refinements (See Supporting Information), which is confirmed by susceptibility and pulsed high magnetic field measurements (See below).

For comparison, we have synthesized three samples with Zn contents of 0, 0.6 and 1.0. All the samples are good insulators with room temperature resistances higher than 20 MΩ. The dc susceptibility measurements are shown in Fig.2(a). With the increasing substitution of Cu4 by Zn, the Weiss temperatures $\theta_W$ are gradually reduced from -100 K, -90 K, to -79 K. The Currie constants, which measure the amount of spins, exhibit a similar trend. The magnetizations under zero field cooling (ZFC) and a field cooling of 100 Oe are shown in Fig.2(b). A splitting between FC and ZFC magnetizations in $Cu_4(OH)_6SO_4$ occurs around $T_c$ ~ 7.5 K indicating a long-range Neel transition[27] and a moderate degree of frustration $f = \theta_W/T_c$ ~ 13 for a triangular lattice[5]. In $Zn_{0.6}Cu_{3.4}(OH)_6SO_4$, the transition temperature is suppressed to 3.5 K and the degree of spin frustration is effectively pushed up to ~ 26. The splitting in the partially substituted compound could be an AF or a spin-glass transition[31]. The splitting completely disappears in $ZnCu_3(OH)_6SO_4$. Further magnetization and heat capacity measurements (see below) indicate that no magnetic transition is found even down to 50 mK (f > 1580). As expected, the successful non-magnetic substitution, which reduces the magnetic coordination number from 6 (triangle lattices in parent compound) to 4 (kagome lattices in the new compound), results in a stronger spin frustration.

We further performed extended magnetic measurements on $ZnCu_3(OH)_6SO_4$. Electron spin resonance (ESR) derivative spectrum of $ZnCu_3(OH)_6SO_4$ at 1.8 K is shown in Fig.3(a). The derivative Lorentzian fittings give an average g factor of ~ 2.19. The dc magnetization down to 0.5 K under 1000 Oe is shown in Fig.3(b). The magnetization upturn at low temperatures is caused by the small mixing of Zn and Cu on kagome planes as observed in other inorganic SL candidates. Using the g factor obtained from ESR measurements, we can estimate an effective amount (~ 9%) of quasi-free defects by fitting susceptibilities from 6~15K, which gives a Weiss temperature, $\theta_D$ ~ -1.16 K (inset of Fig.3(b)). The similar level of quasi-free defects and $\theta_D$ were also reported in herbertsmithites[32-34]. By subtracting the above contributions from quasi-free spins, we obtained bulk susceptibilities dominantly contributed by frustrated (or intrinsic) spins. It allows us to make a simulation of high temperature series expansion (HTSE)[35] for the high temperature part (> 150 K) of bulk susceptibilities. The simulation gives an average antiferromagnetic coupling J ~ 67 K, and an amount of frustrated spins, 2.68 per $ZnCu_3(OH)_6SO_4$[14,32,33]. As cooling down, bulk susceptibilities gradually deviate from the HTSE curve and a broad hump develops around $J/k_B$, suggesting a short-range magnetic correlation on frustrated kagome planes. From 6 to 15 K, bulk susceptibilities are nearly temperature-independent with $\chi_1$ ~ 0.068 $cm^3mol^{-1}$ (Fig.3(b) and (c)). Bulk susceptibilities quickly rise with further cooling down. A blurred but still visible maximum $\chi_2$ ~ 0.27 $cm^3mol^{-1}$ is reached in the range of 0.5 ~ 0.6 K (Fig.3(c)). The observations are quite surprising in such an insulating antiferromagnet without any magnetic ordering. Their important physical meaning will be more clear later in combination with heat capacity measurements. The magnetization under high pulse field (up to 42 T) at 4.2 K is shown in Fig.3(d). The saturation magnetization of defect spins appears below 20 T. Above 20 T, a linear field-dependent magnetization is observed, which slope is $\chi_{up}$ (4.2 K) ~ 0.08 $cm^3mol^{-1}$. The non-saturated intrinsic kagome susceptibility is exactly consistent with the bulk one in Fig. 3(c). After subtracting the linear magnetization ($\chi_{up}H$), the saturation magnetization can be clearly seen, which gives an amount of ~ 8 % defects, in agreement with the above Curie-Weiss fitting. A simple S=1/2 Brillouin function can well describe the saturation magnetization of defects except for a small overshooting at low fields, suggesting that defect spins are somehow coupled rather than completely free[15,34].

Heat capacity experiment is one of the key techniques to probe low-energy excitations. We performed heat capacity measurements down to 50 mK with a dilution refrigerator system. No sharp peak structure is observed (Fig.4 (a) and (d)), suggesting the absence of long-range spin ordering. A broad peak is observed at several Kelvins and shifts to higher temperatures when applying magnetic fields. It is a typical Schottky anomaly arising from magnetic defects as found in some SL candidates[14-16,36]. As shown in Fig.4 (b), the difference between the data under zero and finite fields can be well modeled by Zeeman splitting for defect spins. The Schottky term can be written as $f_d \cdot [C_{sch}(\Delta_{H1}) - C_{sch}(\Delta_{H2})]/T$, where $C_{sch}(\Delta_H)$ is the heat capacity from an $S = 1/2$ spin with an energy splitting $\Delta_H$, and $f_d$ is the fraction of doublets per mol $ZnCu_3(OH)_6SO_4$. In the limit of free spins, $\Delta_H$ should exactly follow the Zeeman splitting with g ~ 2.19 obtained from ESR. The slight discrepancy shown in the inset of Fig.4 (b), may be due to a small coupling between quasi-free spins as suggested in our high magnetic field measurements.

In addition, there is a very small Schottky contribution from hydrogen nuclear spins below 80 mK, which can be seen in the inset of Fig.4 (c) and Fig. 5. The term is centered below 50 mK and its high-energy wing can be exactly simulated by $AT^{-2}$, where A ~ 6.6(2)×$10^{-2}$ mJ·K·$mol^{-1}$, in good accord with $\kappa$-(BEDT-TTF)$_2$Cu$_2$(CN)$_3$ and EtMe$_3$Sb[Pd(dmit)$_2$]$_2$[12,13].

The remaining specific heat from 6 to 15 K follows a form of $\beta T^3 + \gamma_1 T$ after the Schottky contributions from defect spins are removed, as shown in the plot of $C_p T^{-1}$ versus $T^2$ in Fig. 4(c), where $\gamma_1$ ~ 177 mJ·$K^{-2}$·$mol^{-1}$ and $\beta$ ~ 1.18 mJ·$K^{-4}$·$mol^{-1}$ [12,13]. Clearly the term $\beta T^3$ arises from lattice contributions. In order to confirm this and make further quantitative analyses strictly, we have measured the specific heat up to 60 K in both $ZnCu_3(OH)_6SO_4$ and $Zn_{0.6}Cu_{3.4}(OH)_6SO_4$ (Fig.4 (d)). Due to the structural similarity, their specific heats merge together at high temperatures when lattice contributions become dominant and magnetic specific heat is negligible at T > 45 K. So we can applied the strict Debye function with a Debye temperature ~ 225 K, to fit the lattice specific heat at T > 45 K (Fig.4(d)). Below 20 K, the discrepancy between Debye fitting and $T^3$-law is less than 1%, which means that we can safely use $\beta T^3$ to fit lattice specific heat at low temperatures (<15 K).

Intrinsic magnetic specific heat from kagome spins is obtained after further subtracting lattice specific heat, as shown in Fig.4(e). Interestingly, two linear behaviors are found with $\gamma_1$ ~ 177 mJ·K$^{-2}$·mol$^{-1}$ from 6 ~ 15 K and $\gamma_2$ ~ 405 mJ·K$^{-2}$·mol$^{-1}$ below 0.6 K. We further calculated the intrinsic magnetic entropy increase (Fig. 4(f)). Correspondingly it shows two linear regions and saturates at ~ 40 K, which implies that a ground state with macroscopic degeneracy (~ 63 % residual entropy) is realized in the new compound.

In order to further uncover the underlying physics, we put intrinsic bulk susceptibilities and specific heat together in Fig. 5. The temperature regions are highly consistent with each other for both quantities. One may ask if thermal dynamics could play a role in the temperature range of 6~15 K. We indeed tried the fittings for specific heat with a form of $aT^2+\beta T^3$ in the temperature range. But it fails to follow the experimental data. And the simultaneous linear behavior and temperature-independent susceptibility are hard to be understood in term of thermal dynamics. Gapless QSL provides a natural explanation[3,4]. The relative temperature range (0.076 ~ 0.19$\theta_W$) also seems reasonable for a gapless QSL. In the S=1/2 KHA [NH$_4$]$_2$[C$_7$H$_{14}$N][V$_7$O$_6$F$_{18}$], the linear specific heat extends to 5~10 K (~0.17$\theta_W$)[15]. In the S=1 triangular antiferromagnet, Ba$_3$NiSb$_2$O$_9$ (6H-B)[14], it can go up to 7 K(0.093$\theta_W$). Due to a smaller spin moment and stronger frustration configuration, for a S=1/2 HKA one can reasonably expect a higher temperature limit, below which gapless QSL overcomes thermal dynamics.

The linear magnetic specific heat suggests the existence of two gapless QSL states in the material[12-16]. Quantitatively, $\gamma$ and $\chi$ in the lower-temperature QSL state are several times larger than those in the higher temperature QSL state. This suggests that the lower temperature QSL state has a larger density of low-energy states than that of higher temperature one. It seems incompatible with valence-bond crystal (VBC)[10], the gapped Z$_2$[6] and gapless U(1) Dirac[7,8] (where C~T$^2$ and $\chi$~T) SL ground states of HKA, but compatible with the RVB QSL with a "pseudo-Fermi surface"[3,4].

What drives the crossover from the higher temperature QSL state to the stable one at lower temperatures? As mentioned above, the magnetic coupling between well-separated neighboring kagome planes, if exists, should be very weak. So it is less possible that the crossover is related to the change of dimension. Moreover, the linear behaviors in specific heat both above and below the crossover, is also hard to be interpreted by the change of dimension[14]. Most likely, the crossover is driven by spin-orbit coupling. The importance of spin-orbit coupling can be quantitatively indicated by Wilson ratio, R$_W$ = 4$\pi^2$k$_B^2\chi$/(3g$^2\mu_0\mu_B^2\gamma$) [5], which are R$_{W1}$ ~ 1.9 and R$_{W2}$ ~ 3.2 in the higher and lower QSL states respectively. Both ratios are much less than those of most inorganic SL candidates but close to organic SL candidates[5], where spin-orbit coupling is considered to be negligible. The remarkable increase of $R_W$ in the lower temperature gapless QSL state means an enhanced importance of spin-orbit coupling, due to the suppression of thermal fluctuations. From the point of view of microstructures, it is also reasonable as corrugated distorted kagome planes can induce a finite DM interaction.

In conclusion, we synthesized a new QSL candidate, $ZnCu_3(OH)_6SO_4$, which has Cu-1/2 corrugated distorted but well magnetically separated kagome planes. No magnetic ordering is observed even down to 50 mK ( f > 1580). In the temperature ranges of 6 ~15 K and below 0.6 K, intrinsic magnetic specific heat shows a linear temperature dependence and bulk susceptibilities are constants. The facts lead us to an unexpected conclusion that the gapless QSL state at higher temperatures (6 ~ 15 K) re-enters into a QSL state with a larger density of low-energy states at T < 0.6 K, after a crossover possibly driven by spin-orbit coupling. The observations cannot be interpreted by the existing theoretical picture. The present work may lead to a new insight into QSL ground state and its low-energy excitations.

**Methods**

$ZnCu_3(OH)_6SO_4$ (#**1**), $Zn_{0.6}Cu_{3.4}(OH)_6SO_4$ (#**2**) and $Cu_4(OH)_6SO_4$ (#**3**) powder samples were prepared by hydrothermal synthesis from an aqueous suspension obtained from $CuSO_4\ 5H_2O$ (#**1**: 1388 mg, 5.56 mmol. #**2**: 1198 mg, 4.8 mmol. #**3**: 1598 mg, 6.4 mmol), $ZnSO_4\ 7H_2O$ (#**1**: 799 mg, 2.78 mmol. #**2**: 460 mg, 1.6 mmol.

#**3**: 0 mg, 0 mmol) and NaOH (#**1**: 445 mg, 11.12 mmol. #**2**: 384 mg, 9.6 mmol. #**3**: 384 mg, 9.6 mmol) in 34 ml $H_2O$. The aqueous suspension was transferred into a 100 ml Teflon liner, which was capped and placed into a stainless steel pressure vessel. The vessel was heated at 170 $^{o}$C for 4 days, and then cooled down to room temperature at a rate of 0.3 $^{o}$C·min$^{-1}$. A blue-green polycrystalline powder was found at the bottom of the liner, then isolated from the liner by filtration, washed with high purity water, ethyl alcohol and acetone repeatedly and in succession, and dried over by a loft drier at 50 $^{o}$C. By this procedure, 810 mg of $ZnCu_3(OH)_6SO_4$, 700 mg of $Zn_{0.6}Cu_{3.4}(OH)_6SO_4$ and 670 mg of $Cu_4(OH)_6SO_4$ were obtained. The corresponding yields are 96 %, 97 % and 93 % respectively, with respect to the starting material NaOH.

Inductively coupled plasma (ICP) measurements were performed with a Horiba Jobin Yvon Ultima 2 ICP system. The measured ratios of Zn:Cu are 0.91(5) : 3.09 in $ZnCu_3(OH)_6SO_4$ and 0.58(3) : 3.42 in $Zn_{0.6}Cu_{3.4}(OH)_6SO_4$. Scanning electron microscope − X-ray energy dispersive spectrometer (SEM-EDX, JEOL JSM-6700F) measurements were performed and indicated the ratios of Zn:Cu are 1.05(5) : 2.95 in $ZnCu_3(OH)_6SO_4$ and 0.68(5) : 3.32 in $Zn_{0.6}Cu_{3.4}(OH)_6SO_4$. Synchrotron X-ray diffractions (XRD) and X-ray absorption fine structure (XAFS) were performed in the diffraction station (4B9A) of Beijing Synchrotron Radiation Facility (BSRF). The powder samples for XRD were impacted into the 1cm×1cm×1.5mm square tank of slide glasses. The highly monochromatic x-ray from Si (111) double-crystal monochromator was focused on the surface of the powder sample with a 3×1 mm$^2$ light spot. General Structure Analysis System (GSAS) program was used for Rietveld refinements[37]. ESR measurements were performed using a Bruker EMX plus 10/12 CW-spectrometer at X-band frequencies (f ~ 9.36 GHz), equipped with a continuous He gas-flow cryostat. High pulsed field (up to 42 T) magnetization measurements were performed at Wuhan Pulsed High Magnetic Center (WHMFC). T > 2 K magnetization measurements were made with a Quantum Design MPMS system, on ~ 30 mg powder. 2 K > T > 0.5 K magnetization measurements were made with a Quantum Design SQUID with a $^3$He system, on ~ 50 mg powder. T > 2 K specific heats were measured with a Quantum Design PPMS system, on ~ 4 mg dye-pressed pellets. Specific heat between 3.6 K and 50 mK were measured with a Quantum Design PPMS with a dilution refrigerator system, on a 1.7 mg dye-pressed pellet of $ZnCu_3(OH)_6SO_4$. International system of units (SI) is used.

**Acknowledgements**

We thank A. Zorko, H. D. Zhou, R. Yu and T. Li for helpful discussions, Y. G. Cao & Q. F. Huang for assisting with ICP and SEM-EDX experiments. This work was supported by the NSF of China and the Ministry of Science and Technology of China (973 projects: 2011CBA00112 and 2012CB921701). Q.M.Z. was supported by the Fundamental Research Funds for the Central Universities, and the Research Funds of Renmin University of China.

**Figure captions**

**Fig. 1.** (a) Synchrotron X-ray powder diffraction and Rietveld refinement for $ZnCu_3(OH)_6SO_4$ with a incident light of hν = 8639 eV, λ = 1.4352 Å, at room temperature. (b) Vertical view of $ZnCu_3(OH)_6SO_4$ along a axis. All different symmetrical sites and ∠CuO(Cu, Zn) bond angles are labeled. Here sulphur (S) and hydrogen (H) atoms are omitted for simplification. Only hydrogen atoms are omitted in (c), (d) and (e). (c) Polyhedral structure of $ZnCu_3(OH)_6SO_4$ on bc plane. (d) Cu-1/2

kagome network projected on bc plane. Thin green lines show the unit cell. (e) Vertical view of $ZnCu_3(OH)_6SO_4$ along c axis. Cu-1/2 kagome planes are corrugated along *b* axis and well magnetically separated along *a* axis.

**Fig. 2.** Magnetization measurements for $ZnCu_3(OH)_6SO_4$, $Zn_{0.6}Cu_{3.4}(OH)_6SO_4$ and $Cu_4(OH)_6SO_4$ respectively. (a) $HM^{-1}$ versus T data measured under 1000 Oe at 2 ~ 300 K. Colored dash lines are linear (Currie-Weiss) fittings from 150 ~ 300 K, which give Weiss temperatures $\theta_W$ = -79, -90 and -100 K, respectively. (b) $MH^{-1}$ versus T measured under zero-field-cooling (ZFC) and field-cooling (FC, 100 Oe) with a measuring field of 100 Oe at 2 ~ 15 K. For $ZnCu_3(OH)_6SO_4$, the FC and ZFC curves are completely overlapped.

**Fig. 3.** (a) Electron spin resonance (ESR) derivative spectrum of $ZnCu_3(OH)_6SO_4$ measured at 1.8 K. The red curve is the derivative Lorentzian fittings labeled as "der. Lorentz fit". (b) The susceptibilities of $ZnCu_3(OH)_6SO_4$ measured under a field of 1000 Oe at 0.5 ~ 300 K. The red dash curve is the high temperature series expansion (HTSE) simulation of bulk susceptibilities at 150 ~ 300 K. The inset shows the fitting of total susceptibilities including a Curie-Weiss term and a constant bulk term at 5 ~ 16 K, with the adjusted coefficient of determination $R^2$ = 1 and reduced goodness of fit $\chi^2$ ~ $9.5 \times 10^{-9}$. (c) Bulk susceptibilities of $ZnCu_3(OH)_6SO_4$ after subtracting the susceptibilities from defect spins, $\chi_{imp}$. The dot lines are guides to the temperature-independent susceptibilities ($\chi_1$ and $\chi_2$). (d) Magnetization of ZnCu$_3$(OH)$_6$SO$_4$ measured at 4.2 K under high pulsed field. The high pulsed field data (up to 42 T, the black curve) are normalized by PPMS-VSM measurements (up to 14 T, the black circles). The black dot line is a linear fit to the data for $\mu_0H > 20$ T, which slope is considered to be the upper bound of non-saturated intrinsic susceptibilities. The red solid curve and circles labeled as "subtr." are magnetizations from defect spins, obtained by subtracting the above linear contributions from the corresponding total (raw) magnetizations. The blue dash curve labeled as "Brilliouin" is Brillouin function with an amount of ~ 8 % free spins (S=1/2).

**Fig. 4.** (a) Heat capacities of ZnCu$_3$(OH)$_6$SO$_4$ measured under various magnetic fields. Open and closed symbols correspond to PPMS measurements with liquid $^4$He refrigeration and a dilution refrigerator system respectively. (b) The difference between heat capacities under 0 T and $\mu_0H$, [C(0)-C($\mu_0H$)]/T ($\mu_0H$ = 4, 8 and 12 T). The black solid curves are the simulated Schottky anomalies from magnetic defects under 4, 8 and 12 T. Inset: Open squares: energy-level splitting, $\Delta_H$. Solid line: Zeeman splitting for S = 1/2 free spins with g ~ 2.19 obtained from ESR. (c) Heat capacities of ZnCu$_3$(OH)$_6$SO$_4$ after Schottky term subtracted. Dash line: linear fitting for T > 6 K. Inset: The red line is the fitting curve for T < 0.6 K using $C_p = AT^{-2} + \beta T^3 + \gamma_2 T^\alpha$ corresponding to the contributions from nuclear spins, lattice and frustrated spins, where $\beta$ ~ 1.18 mJ·K$^{-4}$·mol$^{-1}$ extracted from T > 6 K data. The fitting gives $\alpha$ ~ 1.03 (1) and A ~ 6.6(2)×10$^{-2}$ mJ·K·mol$^{-1}$. (d) Heat capacities of ZnCu$_3$(OH)$_6$SO$_4$ and Zn$_{0.6}$Cu$_{3.4}$(OH)$_6$SO$_4$. The strict Debye lattice heat capacity exactly follows a T$^3$-law

below 20 K with an error of less than 1%. (e) Heat capacities of $ZnCu_3(OH)_6SO_4$ after Schottky and lattice contributions subtracted. There exist two linear regions: 15 K > T > 6 K and 0.6 K > T > 0.1 K. (f) Intrinsic magnetic entropy increase of $ZnCu_3(OH)_6SO_4$ from the lowest temperature (50 mK). Two dash lines correspond to the linear regions of heat capacities.

**Fig. 5.** Comparison of intrinsic magnetic heat capacities and bulk susceptibilities of $ZnCu_3(OH)_6SO_4$. Both quantities correspond to each other exactly in the two ranges of 6 ~ 15 K and 80 mK ~ 0.6 K. The slight upturn in specific heat below 80 mK is due to nuclear Schottky anomaly as fitted in the inset of Fig.4(c).

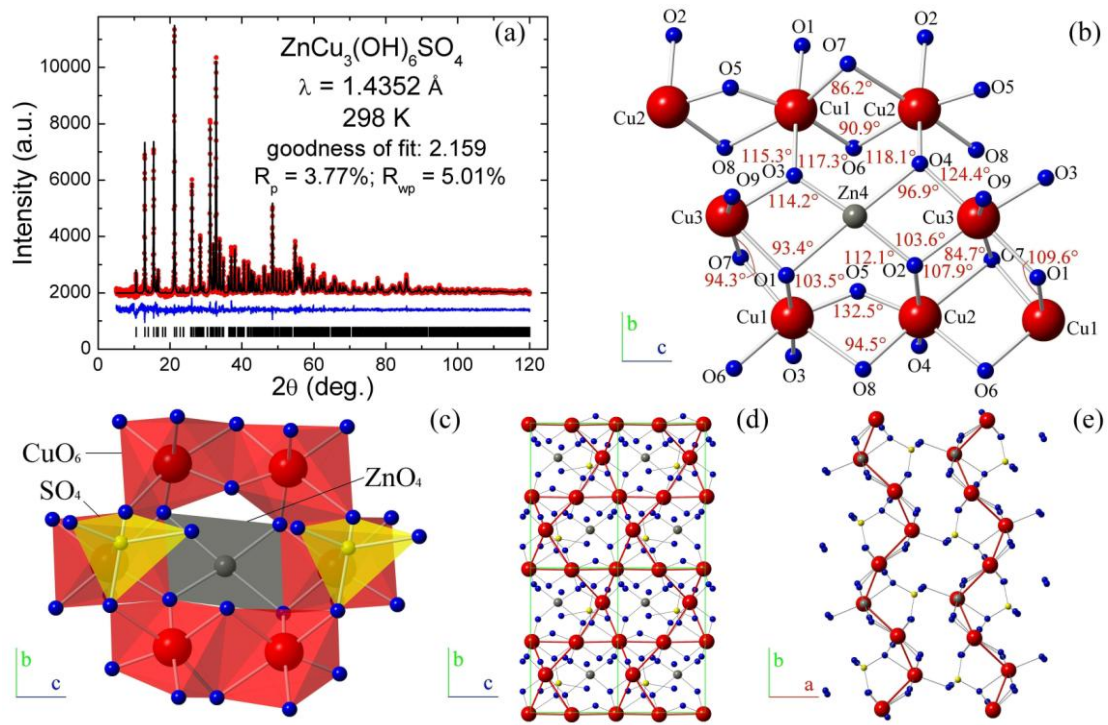

Fig. 1

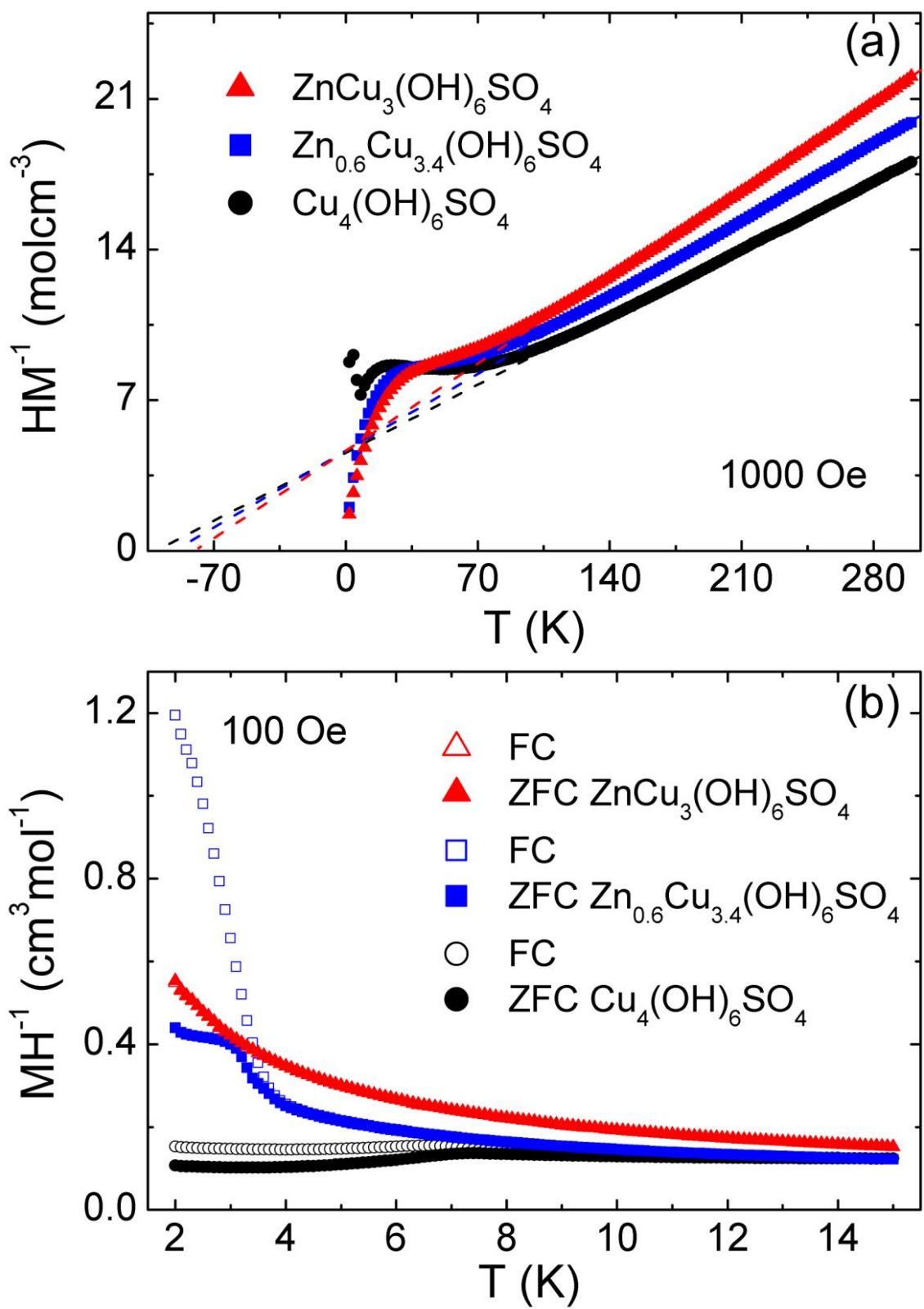

Fig. 2

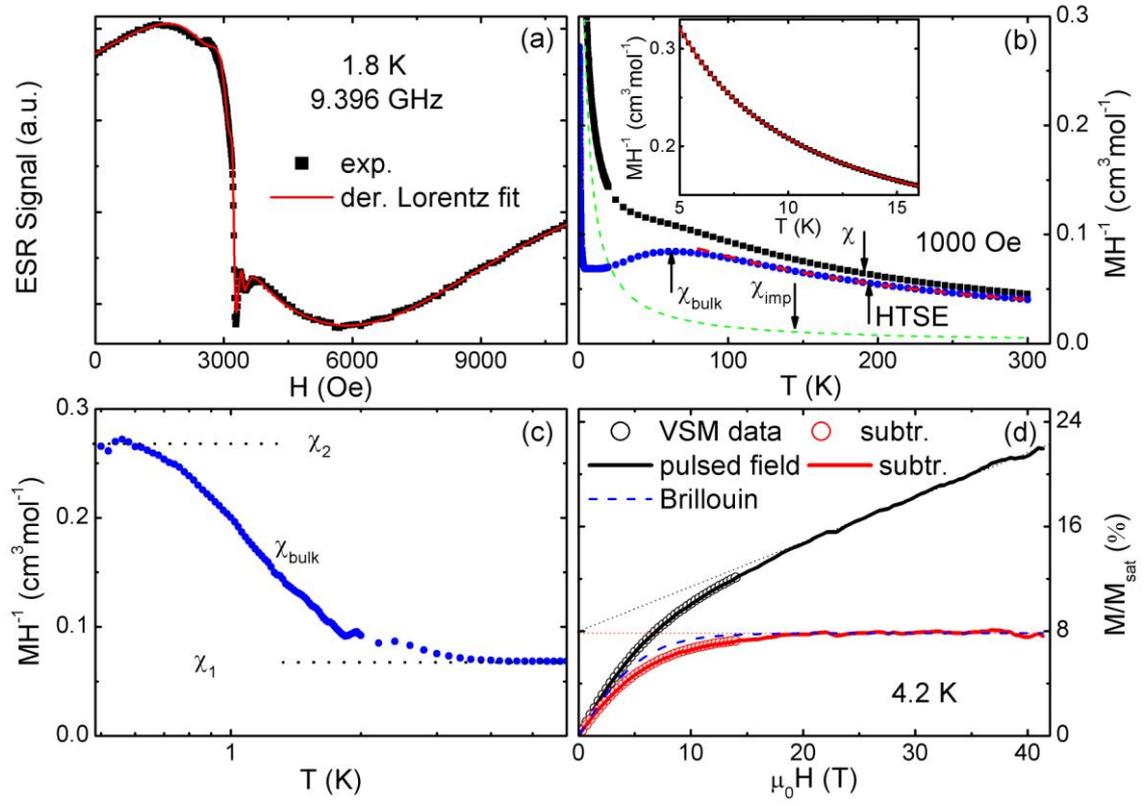

Fig. 3

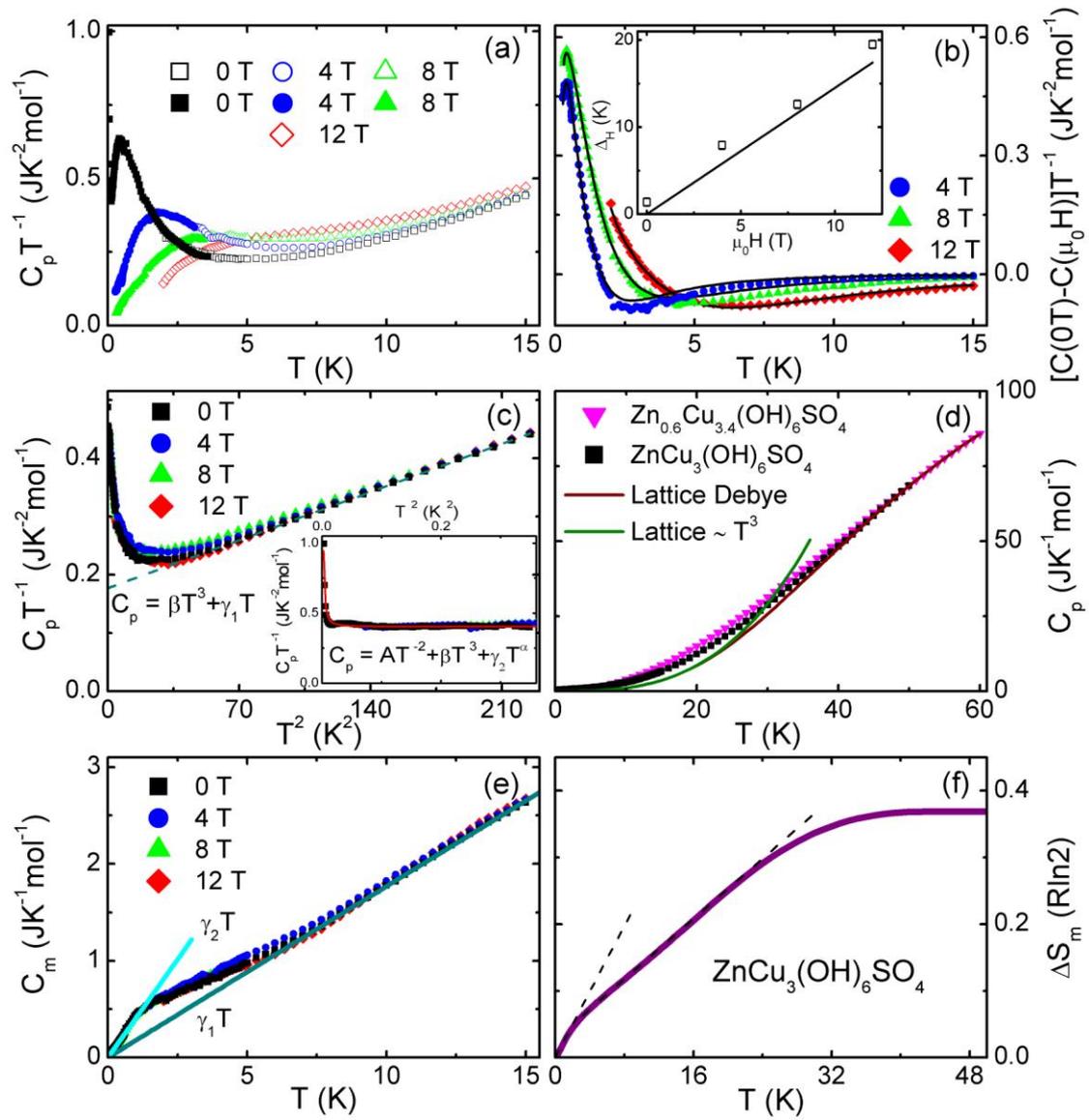

Fig. 4

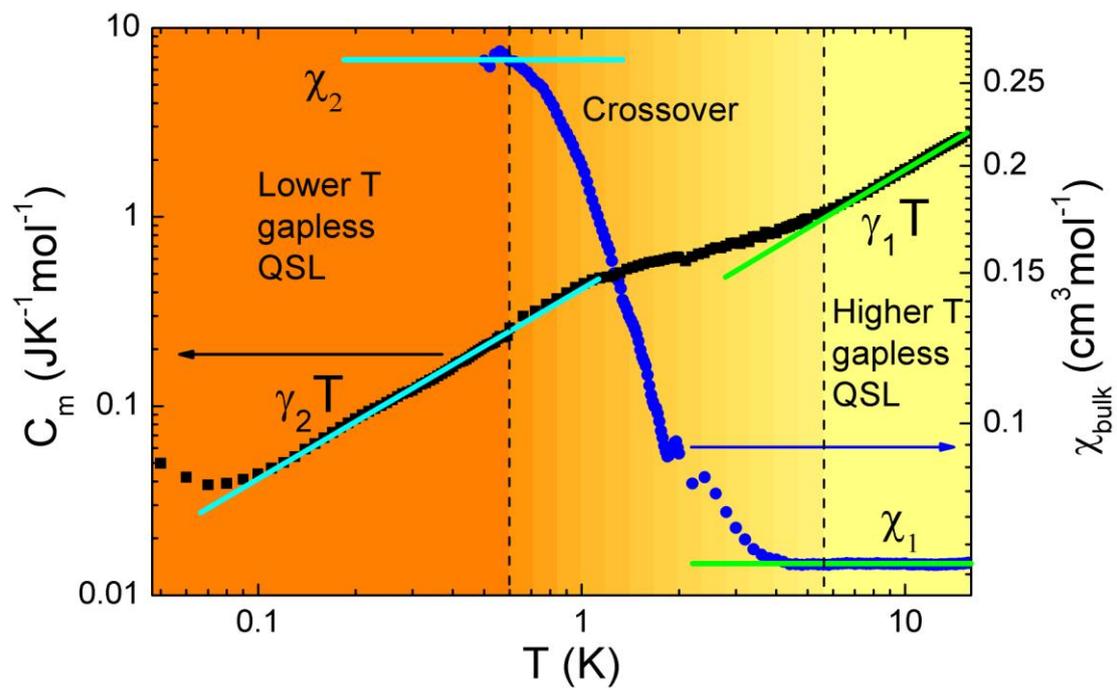

Fig. 5